\documentclass[aps,twocolumn,showpacs,amsmath,amssymb,prl,superscriptaddress, floatfix]{revtex4-1}

\usepackage{graphicx}
\usepackage{dcolumn}
\usepackage{bm}
\usepackage{amssymb}
\usepackage{multirow}

\begin{document}

\title{Formation of density waves via interface conversion of ballistic and diffusive motion}

\date{\today}


\author{Christoph Petri}
\affiliation{Zentrum f\"ur Optische Quantentechnologien, Universit\"at Hamburg, Luruper Chaussee 149, 22761 Hamburg, Germany}%
\author{Florian Lenz}
\affiliation{Zentrum f\"ur Optische Quantentechnologien, Universit\"at Hamburg, Luruper Chaussee 149, 22761 Hamburg, Germany}%
\author{Benno Liebchen}
\affiliation{Zentrum f\"ur Optische Quantentechnologien, Universit\"at Hamburg, Luruper Chaussee 149, 22761 Hamburg, Germany}%
\author{Fotis K. Diakonos}
\affiliation{Department of Physics, University of Athens, GR-15771 Athens, Greece}
\author{Peter Schmelcher}
\email[]{Peter.Schmelcher@physnet.uni-hamburg.de}
\affiliation{Zentrum f\"ur Optische Quantentechnologien, Universit\"at Hamburg, Luruper Chaussee 149, 22761 Hamburg, Germany}%

\begin{abstract}
We develop a mechanism for the controlled conversion of ballistic to diffusive motion and vice versa. This process takes place at the interfaces of domains with different time-dependent forces in lattices of laterally oscillating barrier potentials. As a consequence long-time transient oscillations of the particle density are formed which can be converted to permanent density waves by an appropriate tuning of the driving forces. The proposed mechanism opens the perspective of an engineering of the nonequilibrium dynamics of particles in inhomogeneously driven lattices.
\end{abstract}

\maketitle

\section{Introduction}
Time-dependent forces are often the origin for the appearance of complex dynamics in physical systems \cite{Ott:1992}. Spectacular effects such as the directed transport of particles in the absence of a net force can be evoked by breaking certain space-time symmetries (parity and time-reversal \cite{Flach:2000}) homogeneously, i.e. through the application of a spatially independent driving force. In the literature this phenomenon is usually referred to as the ratchet effect \cite{Reimann:2002} and has attracted a great deal of attention since it is the working principle of molecular \cite{Magnasco:1993,Julicher:1997} or quantum \cite{Ponomarev:2009} motors. In the case of Hamiltonian (or deterministic) ratchets \cite{Mateos:2000,Brito:2001,Marchesoni:2002} both setups with fully chaotic \cite{Monteiro:2002,Hutchings:2004,Brumer:2006} and mixed phase space \cite{Flach:2000,Gong:2004,Casati:2007,Denisov:2001,Denisov:2002,Schanz:2001,Schanz:2005,Dittrich:2000,Petri:2010} have been explored. Experimentally, quantum ratchets have been demonstrated e.g. in Josephson junction arrays \cite{Majer:2003} and for cold atoms loaded into periodically amplitude-modulated optical lattices \cite{Salger:2009}.

In view of the above it is an intriguing perspective to consider the consequences of a breaking of the parity and time-reversal symmetry inhomogeneously by applying a spatially varying driving force. Recently \cite{Liebchen:2011} a phase-modulated driven lattice with unit cells consisting of a few sites has been shown to provide a mechanism which leads to a selective patterned trapping of particles on certain sites. The latter could be useful for atom lithography or as an ingredient for quantum information processing. Here we explore a different driven lattice reminiscent of superlattices in the physics of semiconductor heterostructures but now with a spatiotemporal embedding. It consists of a lattice of domains, each containing many unit cells, and subject to a different time-periodic force. The driving is chosen such that neighboring domains possess oppositely directed currents (``local'' ratchets). Our focus is the analysis of the nonequilibrium dynamics in such bimodally driven devices. At the interfaces of the domains a controlled conversion of the dynamics from diffusive to ballistic behavior and vice versa is possible and can be understood in terms of the overlap of the Poincar\'{e} surfaces of section (PSS) of the individual domains. As a result we observe the formation of transient spatial density oscillations. The latter persist in time if the lattice is fed with a constant current of particles. By switching appropriately the applied forces it is possible to manipulate these density waves.


\begin{figure}
\includegraphics[width=\columnwidth]{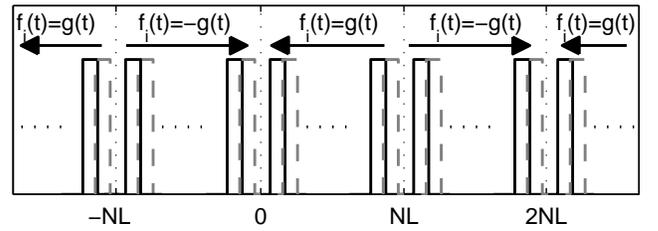}
\caption{\label{fig:fig1} Schematic illustration of the driven lattice. Interfaces between the domains with different driving laws are at the positions $\{0,\pm NL, \pm 2NL\}$. Arrows indicate the direction of the local currents.}
\end{figure}

\section{Driven lattice Hamiltonian}
We consider noninteracting particles in a spatiotemporally driven lattice with potential barriers of equal height $V_0$ and width $l$ described by the Hamiltonian
\begin{equation}\label{eq:ham}
H(x,p,t)=\frac{p^2}{2m}+V_0 \sum_{i=-\infty}^\infty \Theta \left(\frac{l}{2}-\left| x-x_{0,i}-f_i(t) \right| \right),
\end{equation}
where $m$ is the mass of the particles and $x_{0,i}=i\,L$ is the equilibrium position of the $i$th barrier. $f_i(t)$ is a periodic driving law, which depends on the site index $i$ according to $f_i(t)=(-1)^{\lfloor i/N \rfloor}\,g(t)$ with $g(t)=C\left\{\cos\left(\omega\,t\right)+\sin\left(2\omega\,t\right)\right\}$ being a biharmonic function and $\lfloor x \rfloor$ is the largest integer $k$ with $k\leq x$ (see Fig. \ref{fig:fig1}). The lattice is composed of domains each consisting of $N$ laterally oscillating barriers obeying the same driving law and which are connected at corresponding interfaces. Each domain possesses the length $L_{\textrm{B}}=N\cdot L$ ($N=10^4$) and, as shown in Fig. \ref{fig:fig1}, the driving law $f_i(t)$ alternates from one domain to the next. The amplitude $C$ of the driving is fixed such that $-1\leq g(t) \leq1$, i.e. $C\approx 0.57$ and without loss of generality we assume $\omega=1$ and $m=1$, as well as $l=1$, $V_0=2.2$ and $L=5$.

\section{Transport properties of the lattice} Applying the driving law $g(t)$ to the entire lattice leads to directed currents since the relevant spatiotemporal symmetries derived in \cite{Flach:2000} are broken. By integrating over the main chaotic sea with respect to $p=0$ the transport velocity can be estimated to $v_{\textrm{T}}=-0.2649$ \cite{Petri:2010}. In a domain with the driving law $f_i(t)=g(t)$ particles in the chaotic sea are on average transported in negative $x$-direction. Since the driving changes its sign from one domain to the next one, i.e. $f_i(t)=-f_{i+N}(t)$, the direction of the local current alternates, too, and transport is reversed. There are two types of interfaces that repeat in an alternating manner in the lattice. One type is characterized by incoming `convergent' currents from both neighboring domains whereas the second one has `divergent' outgoing currents pointing towards the neighboring interfaces (see Fig. \ref{fig:fig1}). Naturally, one could therefore expect that interfaces with incoming currents show particle accumulation whereas those with outgoing currents exhibit particle depletion. Although each domain shows directed transport and an according drift of the particles, the complete lattice consisting of infinitely many domains does not exhibit directed currents since the equations of motion are invariant under the transformation $T: x \rightarrow -x+NL, \, t \rightarrow t$ \cite{Flach:2000}. Yet, an ensemble of particles reaches this zero current limit of the complete driven lattice only for extremely long time scales by far not reached in the present work.


\begin{figure}
\includegraphics[width=\columnwidth]{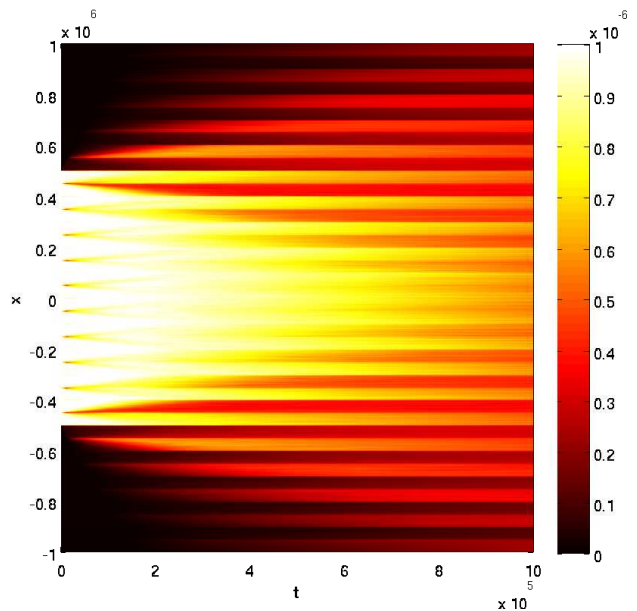}
\caption{\label{fig:fig2} (Color online) Particle density $\rho(x,t)$. The initial ensemble is distributed according to $\rho(x,t=0)=10^{-6}\cdot\Theta(10^5L-|x|)$ and $\rho(v,t=0)=5\cdot\Theta(0.1-|v|)$ ($V_0=2.2$). For $t\gtrsim10^5$ a periodic modulation of the spatial distribution of the particles in the lattice is clearly visible.}
\end{figure}

\section{Results}
We explore the dynamics of an ensemble of $10^6$ particles which are initially uniformly distributed in the lattice occupying 20 domains centered at the origin and which homogeneously fill the chaotic sea at low velocities $|v|\leq 0.1$. (For details of the numerical method see Refs. \cite{Koch:2008,Petri:2010}). The spatiotemporal evolution of the particle density profile (see Fig. \ref{fig:fig2}) shows due to the diffusion of the chaotic trajectories a significant spatial broadening. More importantly, we observe the emergence of a periodic stripe-like modulation of the density, i.e. for a fixed time the spatial distribution of the particles has alternating flat maxima and minima corresponding to the length of one domain. This behavior persists until $t\approx 5\cdot 10^6$ and gradually disappears thereafter. Let us analyze this in some more detail.

\begin{figure}
\includegraphics[width=\columnwidth]{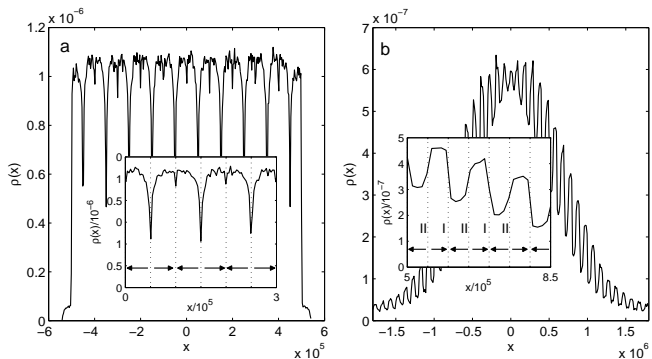}
\caption{\label{fig:fig3} Spatial distribution of the particles $\rho(x)$ for $t=10^4$ (a) and $t=10^6$ (b). The arrows in the magnifications indicate the direction of the local current in the domain.}
\end{figure}

For $t=10^4$ (Fig. \ref{fig:fig3} (a)) the envelope behavior of $\rho(x)$ is still similar to the initial distribution $\rho_0(x)$ but there occur sharp dips at the positions of the interfaces. The dips are significantly more pronounced at the interfaces with outgoing local currents (inset of Fig. \ref{fig:fig3} (a)). At $t=10^6$  ( Fig. \ref{fig:fig3} (b)) the envelope of $\rho(x)$ has broadened and the density modulations are much more pronounced extending over an entire domain of the lattice (inset of Fig. \ref{fig:fig3} (b)). For $x>0$ $\rho(x)$ possesses plateau-like maxima in domains with $v_{\textrm{T}}>0$ and corresponding minima in domains with $v_{\textrm{T}}<0$ and vice versa for $x<0$.

\begin{figure}
\includegraphics[width=\columnwidth]{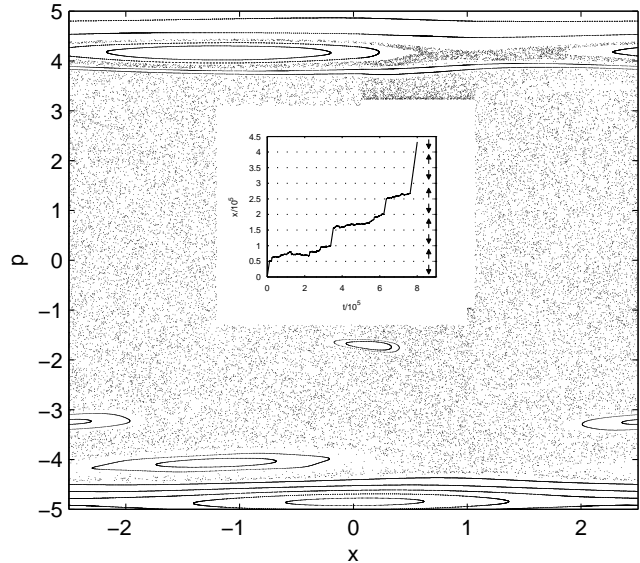}
\caption{\label{fig:fig4} Stroboscopic Poincar\'{e} surface of section (PSS) for a single domain of the lattice with $g(t)=C\left\{\cos\left(\omega\,t\right)+\sin\left(2\omega\,t\right)\right\}$ ($C\approx 0.57$, $V_0=2.2$, $l=1$ , $L=5$). Inset: Individual trajectory. Dotted lines: Positions of the interfaces with small arrows indicating the local currents in the corresponding domain.}
\end{figure}

To understand the mechanism which is responsible for the formation of the density modulation, it is instructive to look at the trajectories of individual particles. Fig. \ref{fig:fig4} (inset) shows the typical dynamics which consists of alternating phases of chaotic and ballistic behavior. Initially, the particle propagates ballistically opposite to the direction of the local particle current until it reaches the next interface where a conversion to chaotic dynamics occurs. In the next domain the trajectory diffuses chaotically with a net drift according to the directed transport in this domain. When it arrives at the second interface a conversion back to ballistic behavior is observed. The dynamical conversion process repeats at each interface until the particle starts a very long flight which lasts over several domains. Typically, we use the term ``ballistic flight'' for time intervals during which the particles travel over a long distance $x\geq10L$ in the same direction.

To understand the interplay between ballistic and diffusive motion at the interfaces it is useful to visualize the dynamics by means of stroboscopic Poincar\'{e} surface of sections (PSS). A super PSS containing the detailed dynamics for times up to $t \gg 10^7$ over the entire lattice possesses an enormously rich structure at any scale and is in practice inaccessible. For the description of the transient dynamics characteristic for times $t \leq 10^6$ the PSS of a domain (PSSD) is more adequate (see Fig. \ref{fig:fig4}). For momenta $|p|\gtrsim4.3$ the dynamics is mostly regular, i.e. in this region the phase space is predominantly foliated by deformed tori of the unperturbed system. A big elliptic island centered around a period one orbit at $x\approx 0, p\approx-4.9$ is encountered. Furthermore, we observe a chaotic sea confined by invariant curves to $3.9\lesssim p\lesssim 4.3$. Trajectories which start in this sea traverse the lattice domain in the direction of their initial velocity. Particles starting in the large chaotic sea ($-4.3\lesssim p\lesssim 3.9$) wander diffusively through their domain whereas trajectories belonging to the embedded regular islands travel only in the direction of their initial momentum. Domains with driving laws $g(t)$ respectively $-g(t)$ possess mirror symmetric PSSDs with respect to $p=0$. At the interfaces between the regions with different driving laws the PSSDs of the two domains are concatenated. By crossing such an interface it is possible that a trajectory is injected from the chaotic sea into a ballistic island. Then the particle stays in this island and traverses the subsequent domain until it reaches the next interface where a conversion back to the chaotic sea can occur. Trajectories (inset of Fig. \ref{fig:fig4}) thus exhibit alternating ballistic and diffusive phases each on the scale of a single domain.

Let us now return to the discussion of the evolution of $\rho(x)$. In the beginning ($t=10^4$) we observe a sharp depletion of the particle density at the positions of the interfaces (Fig. \ref{fig:fig3} (a)). Every time a chaotic trajectory crosses an interface it is possible that a ballistic flight starts at this position. By analyzing the spatial distribution of the starting positions of ballistic flights $\rho(x_{\textrm{start}})$ we find narrow peaks localized at the interfaces. Apparently, this process leads to a fast local reduction of the spatial distribution of the particles $\rho(x)$ at these positions since compared to diffusive trajectories, the ballistic particles travel with a high velocity away from the interface. However, if the local currents are incoming this effect competes with the accumulation of particles obeying chaotic dynamics. Therefore, the dip is more pronounced for interfaces with outgoing local currents.

Let us now develop an understanding of the longer-time transient behavior leading to the plateau-like density modulations (Fig \ref{fig:fig3} (b)). The initial ensemble is localized in the chaotic sea of the phase space of the lattice. In every domain the particles experience the corresponding drift velocity and accumulate at the interfaces I with incoming directed currents. Even if a `deeper' reinjection from the interface into an adjacent domain again occurs (which is a rare event) it will return to this interface at a somewhat later time. Hence, chaotic trajectories are transported several times back to the interface until a conversion from chaotic to ballistic dynamics occurs. Accordingly, diffusive particles are localized in a small spatial region around these interfaces I until their position in phase space coincides with a regular structure in the phase space of the adjacent domain which leads to a conversion into regular ballistic motion. When such a ballistic particle reaches the next interface II (outgoing directed currents) the trajectory can either remain ballistic and traverse the subsequent domain in a comparatively short time or the particle is injected from the ballistic island back into the chaotic sea. It is however important to keep track of the fact that the ballistic trajectories cross the interface II and typically enter up to approximately $10^3$ unit cells into the next domain where they acquire a diffusive character. In the latter domain the particles experience a drift to the adjacent interface I, since the local current points away from the interface II. Obviously, this process does not occur at the interface I with exclusively incoming currents. Note that domains with left pointing current can be traversed to the right only by ballistic trajectories. As a result an enhanced number of diffusive trajectories occurs in domains with right pointing current for $x>0$ and vice versa. Since the dwell time of chaotic trajectories in a domain is approximately one order of magnitude larger than the dwell time of ballistically travelling particles this results in a periodic modulation of $\rho(x)$. It is crucial to note that the Hamiltonian itself does not reflect such an asymmetry with respect to the domains with left and right-pointing currents corresponding to driving laws $g(t)$ respectively $-g(t)$. Indeed, the asymptotic steady state of the time-evolved ensemble ($t\gg10^7$) possesses a uniform density.

\begin{figure}
\includegraphics[width=\columnwidth]{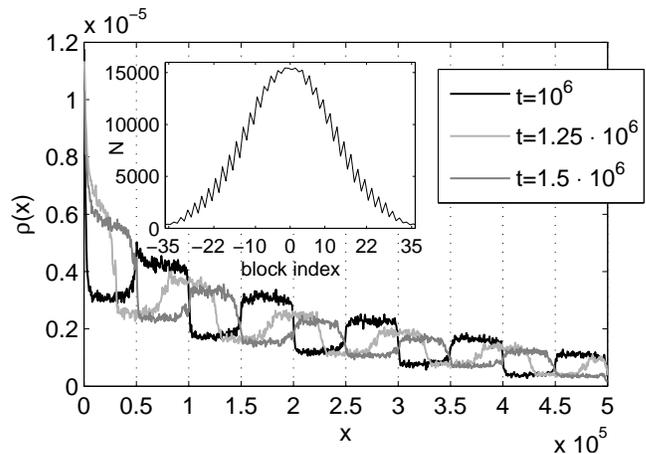}
\caption{\label{fig:fig5} Snapshots of the density $\rho(x)$ in a finite, open system with a constant current of particles entering the lattice. Dotted lines: Positions of the interfaces at $t=10^6$ and $t=1.5 \cdot 10^6$ are identical. Inset: Spatial distribution of particles obtained by modelling via the Markov chain.}
\end{figure}

The above nonequilibrium dynamics can be described by a Markov chain that represents an ``interface mapping'' comprising the dynamics within a domain. Components of the Markov state are the dynamical character of a trajectory (chaotic / ballistic), the direction of motion (left / right), the position of the particle and the point in time it reaches the current interface. The conditional transition probabilities $p_{ij}$ of a dynamical conversion at an interface, which constitute the transition matrices can be obtained by calculating the overlap of the different regular and chaotic structures in phase space and augmenting it by local characteristics of the different interfaces. The inset of Fig. \ref{fig:fig5} (compare to Fig. \ref{fig:fig3} (b)) demonstrates that the Markov chain indeed reproduces the long-term transient spatial distribution of the particles in the lattice.

It is possible to manipulate the above density profile such that it becomes a steady state of the spatiotemporally driven lattice. To this end, we consider a finite open system composed of 10 domains with an incoming constant current of particles. After a transient phase the spatial distribution of the particles approaches a steady state (Fig. \ref{fig:fig5}) with an appearance similar to the case shown in Fig. \ref{fig:fig3} (b). It is now possible to render this steady state into a time propagating density wave by generating a movement of the interfaces through the lattice. The latter is achieved by a consecutive switching of the driving laws of the barriers. We then observe that the spatial distribution of the particles follows this change of the lattice adiabatically, i.e. the appearance of $\rho(x,t)$ is preserved during this process resulting in a propagating density wave.


\section{Conclusion}
We have demonstrated novel nonequilibrium phenomena in 1D driven lattices associated with the inhomogeneous breaking of parity and time-reversal symmetries. Dynamical conversions of trajectories from ballistic to diffusive motion are observed at the interfaces between differently transporting domains in the lattice. For a beam of particles entering the device this mechanism leads to a density modulation which can be manipulated by switching the driving laws of the individual barriers. A suitable experimental setup which could demonstrate such density waves are multiple layered semiconductor heterostructures driven according to locally different AC-voltages or applied laser fields. Cold atoms in optical lattices formed by counterpropagating laser beams are an alternative experimental setup to verify our findings. By mounting the mirrors on piezoelectric crystals a laterally oscillating one-dimensional potential can be created.


\acknowledgments
This work has been performed within the Excellence Cluster Frontiers in Quantum Photon science, which is supported by the Joachim Herz Stiftung. Financial support by the DAAD in the framework of an exchange program with Greece (IKYDA) is acknowledged.

\end{document}